\RequirePackage{filecontents}

\RequirePackage{snapshot}
\documentclass[aps,pra,reprint,showpacs,secnumarabic,groupedaddress,floatfix,longbibliography]{preprint}

\renewcommand\documentclass[2][]{}
\let\origparagraph=\paragraph
\AtBeginDocument{%
  \let\paragraph=\origparagraph
}
\makeatother
\documentclass[reprint,aps,pra,showpacs,floatfix,groupedaddress,balancelastpage,longbibliography]{revtex4-1}

\makeatletter
\def\@bibdataout@aps{%
 \immediate\write\@bibdataout{%
  @CONTROL{%
   apsrev41Control%
   \longbibliography@sw{%
    ,author="08",editor="1",pages="1",title="0",year="0"%
   }{%
    ,author="08",editor="1",pages="0",title="",year="1"%
   }%
  }%
 }%
 \if@filesw
  \immediate\write\@auxout{\string\citation{apsrev41Control}}%
 \fi
}%
\makeatother

\usepackage[T1]{fontenc}
\usepackage{amsmath}
\usepackage{txfonts}
\usepackage{tgtermes}
\usepackage[altg]{qtxmath}
\renewcommand*{\vec}[1]{\boldsymbol{#1}}
\usepackage{microtype}
\usepackage{graphicx}
\graphicspath{{figures/}}
\usepackage[breaklinks=true,
pdfauthor={Nicolas Teeny, Christoph H. Keitel and Heiko Bauke},
pdftitle={Virtual-detector approach to tunnel ionization and tunneling times}]{hyperref}

\newcommand*{\D}{\mathrm{d}}

\newcommand*{\I}{\mathrm{i}}
\newcommand*{\op}[1]{\hat{#1}}

\DeclareMathOperator*{\argmax}{arg\,max}
\let\Re\relax

\DeclareMathOperator{\Re}{Re}

\begin{document}

\title{Virtual-detector approach to tunnel ionization and tunneling times}

\author{Nicolas Teeny}
\email{nicolas.teeny@mpi-hd.mpg.de}
\affiliation{Max-Planck-Institut f{\"u}r Kernphysik, Saupfercheckweg~1, 69117~Heidelberg, Germany}
\author{Christoph H. Keitel}
\affiliation{Max-Planck-Institut f{\"u}r Kernphysik, Saupfercheckweg~1, 69117~Heidelberg, Germany}
\author{Heiko Bauke}
\email{heiko.bauke@mpi-hd.mpg.de}
\affiliation{Max-Planck-Institut f{\"u}r Kernphysik, Saupfercheckweg~1, 69117~Heidelberg, Germany}

\begin{abstract}
  Tunneling times in atomic ionization are studied theoretically by a
  virtual detector approach.  A virtual detector is a hypothetical
  device that allows one to monitor the wave function's density with
  spatial and temporal resolution during the ionization process.  With
  this theoretical approach, it becomes possible to define unique
  moments when the electron enters and leaves with highest probability
  the classically forbidden region from first principles and a tunneling
  time can be specified unambiguously.  It is shown that neither the
  moment when the electron enters the tunneling barrier nor when it
  leaves the tunneling barrier coincides with the moment when the
  external electric field reaches its maximum.  Under the tunneling
  barrier as well as at the exit the electron has a nonzero velocity in
  the electric field direction.  This nonzero exit velocity has to be
  incorporated when the free motion of the electron is modeled by
  classical equations of motion.
\end{abstract}


\maketitle

\section{Introduction}
\label{sec:intro}

In a seminal work, MacColl \cite{MacColl:1932:Note_on} studied the time
that may be associated with the process of a particle approaching from
far away a potential barrier of a height larger than the particle's
energy and eventually tunneling through the barrier.  A series of
subsequent works led to various definitions of tunneling times and
different physical interpretations
\cite{Hauge:Stvneng:1989:Tunneling_times,Landauer:Martin:1994:Barrier_interaction,Maji:Mondal:etal:2007:Tunnelling_time}.
Also many efforts have been directed towards measuring
\cite{Steinberg:Kwiat:etal:1993:Measurement_of,Gueret:Marclay:Meier:1988,Mugnai:Ranfagni:etal:2000:Observation_of,Enders:Nimtz:1993:Evanescent-mode_propagation,Winful:2006:Tunneling_time,Patino:Kelkar:2015:Experimental_determination}
tunneling times.  

\begin{figure}
  \centering
  \includegraphics[scale=0.45]{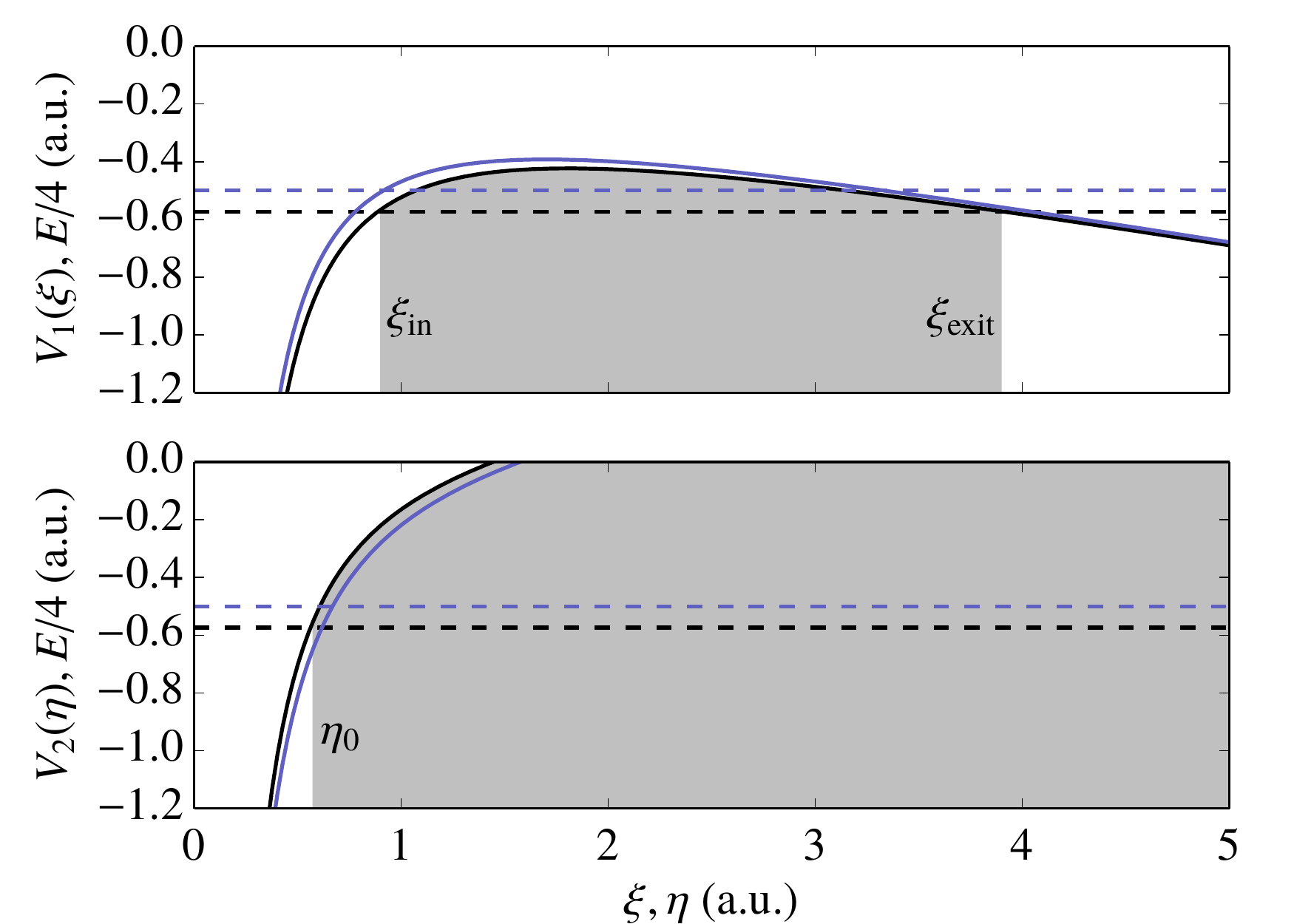}
  \caption{Tunneling potentials $V_1(\xi)$ and $V_2(\eta)$ (solid lines)
    and the ground state energy level $E$ (dashed lines) for the
    two-dimensional Coulomb problem with (black) and without (light
    blue) accounting for the Stark effect in parabolic coordinates $\xi$
    and $\eta$ for $\mathcal{E}=1\,\mathrm{a.u.}$  Intersections
    between the potential lines and the energy level determine the
    borders between classically allowed classically forbidden regions,
    i.\,e., the tunneling entry $\xi_\mathrm{in}$, tunneling exit
    $\xi_\mathrm{exit}$, and $\eta_0$.  See the
    \hyperref[sec:appendix]{Appendix} for details.}
  \label{fig:tunneling_potentials}
\end{figure}

A related still open and actively studied problem of atomic physics is
the question how long it takes to ionize an atom via an electron
tunneling through the potential barrier
\cite{Keldysh:1965:Ionization,Uiberacker:Uphues:etal:2007:Attosecond_real-time,Ban:Sherman:etal:2010:Time_scales,Eckle:Pfeiffer:etal:2008:Attosecond_Ionization,Eckle:Smolarski:etal:2008:Attosecond_angular,Kheifets:Ivanov:2010:Delay_in,Klaiber:Yakaboylu:etal:2013:Under-the-Barrier_Dynamics,Zhao:Lein:2013:Determination_of,McDonald:Orlando:etal:2013:Tunnel_Ionization,Yakaboylu:Klaiber:etal:2013:Relativistic_features,Yakaboylu:Klaiber:etal:2014:Wigner_time,Orlando:McDonald:etal:2014:Tunnelling_time,Landsman:Weger:etal:2014:Ultrafast_resolution,Klaiber:Hatsagortsyan:etal:2015:Tunneling_Dynamics,Torlina:Morales:etal:2015:Interpreting_attoclock,Ivanov:Kim:2015:Relativistic_approach,Landsman:Keller:2015:Attosecond_science,Teeny:etal:2016:Ionization_Time}
which is formed by the electron's binding energy and the Coulomb
potential bent by the time-dependent electric field's potential; see
Fig.~\ref{fig:tunneling_potentials}.  Employing the angular streaking
technique the tunneling dynamics can be studied on subcycle time scales
\cite{Eckle:Pfeiffer:etal:2008:Attosecond_Ionization,Eckle:Smolarski:etal:2008:Attosecond_angular}
in the so-called attoclock experiments, aiming to determine when the
electron ionizes from a bound state.  The attoclock experiments
stimulated renewed efforts towards defining a well-founded tunnel
ionization time and determining the tunnel ionization time.  A consensus
on a suitable 
definition of tunneling time and the interpretation of experimental
results is, however, still lacking
\cite{Maquet:Caillat:etal:2014:Attosecond_delays}.

In the case of tunnel ionization the experimental determination of a
tunneling time is complicated by the fact that it is notoriously
difficult to determine the starting (and ending) moment of the tunnel
dynamics.  There is no apparent reason to assume that the electron
enters the barrier at the instant of the electric field maximum.  In
fact, we are going to demonstrate that one has to carefully distinguish
between the tunneling time, i.\,e., the time to cross the tunneling
barrier, and the tunneling delay, i.\,e., the time delay until the
electron becomes quasi-free with respect to some external reference,
e.\,g., the moment of maximal electric field strength.

Experimentally the quantum dynamics in the vicinity of the tunneling
barrier cannot be studied directly, i.\,e., it is not possible to place
a detector close to the atomic tunneling barrier.  Thus, information
about the tunneling dynamics has to be inferred from measurable
asymptotic quantities, e.\,g., the momentum distribution of the photo
ionized electrons.  In attoclock experiments, an electron is ionized by
an elliptically polarized few-cycle pulse. This quasi-free electron is
accelerated in the rotating electric field, and in this way the instant
of ionization $t_\mathrm{exit}$ is mapped to the final angle of the
momentum vector in the polarization plane.  Ignoring Coulomb effects as
well as non-dipole effects, the electron's final momentum
$\vec p(t_\mathrm{f})$ and the instant of ionization $t_\mathrm{exit}$
are related according to the two-step model
\cite{Gallagher:1988:Above-Threshold_Ionization,Corkum:Burnett:etal:1989:Above-threshold_ionization,Corkum:1993:Plasma_perspective,Schafer:Yang:etal:1993:Above_threshold}
via
\begin{equation}
  \label{eq:simple}
  \vec p(t_\mathrm{f}) = 
  \vec p(t_\mathrm{exit}) +
  q \int_{t_\mathrm{exit}}^{t_\mathrm{f}}\vec E(t)\,\D t\,,
\end{equation}
where $\vec E(t)$ denotes the electric field, which vanishes for
\mbox{$t\to t_\mathrm{f}$}, $q$ denotes the electron's charge, and
$\vec p(t_\mathrm{exit})$ is its initial momentum, which is usually
assumed to be zero.  Assuming that the ionized electron becomes free at
the instant of the electric field maximum $t_0$, the formula
\eqref{eq:simple} predicts a final momentum with a specific direction
\cite{Landsman:Keller:2015:Attosecond_science}.  Deviations of an
experimentally observed final momentum direction from this prediction
may be due to a delay between the instant of maximal field strength and
the instant of ionization, i.\,e., $t_0\ne t_\mathrm{exit}$.  But also
Coulomb effects \cite{Kaushal:Smirnova:2013:Nonadiabatic_Coulomb}, a
nonzero initial momentum $\vec p(t_\mathrm{exit})$
\cite{Teeny:etal:2016:Ionization_Time}, nondipole effects, or quantum
effects, which may play a role for the dynamics at the vicinity of the
tunneling exit, cause deviations from the simple two-step model
\eqref{eq:simple}.  Thus, the interpretation of attoclock experiments
requires a precise model of the electron's motion from the barrier exit
to the detector.  The value of a possible tunneling delay depends
crucially on the theoretical model
\cite{Torlina:Morales:etal:2015:Interpreting_attoclock}, which is
employed to calibrate the attoclock.

In view of the difficulties of determining tunneling times from
asymptotic momentum measurements, we consider a complementary
theoretical approach based on \emph{ab initio} solutions of the
time-dependent Schr{\"o}dinger equation and virtual detectors
\cite{Feuerstein:Thumm:2003:On_computation,Wang:Tian:etal:2013:Extended_Virtual},
which allows us to link tunneling times to observables related to the
quantum dynamics in the vicinity of the tunneling barrier.  This paper
is organized as follows: To keep the presentation self-contained, the
concept of a virtual detector is summarized in
Sec.~\ref{sec:virtual_detectors}.  In Sec.~\ref{sec:results} numerical
results for tunneling times as determined by the virtual detector
approach are presented and discussed.  Mainly, we answer the questions:
When does the electron enter the barrier and exit the tunneling barrier;
and, accordingly, how much time does the electron spend under the
barrier?  Additionally, we compare the quantum trajectory of the electron
to the one predicted by the two-step model
\cite{Gallagher:1988:Above-Threshold_Ionization,Corkum:Burnett:etal:1989:Above-threshold_ionization,Corkum:1993:Plasma_perspective,Schafer:Yang:etal:1993:Above_threshold}. Finally,
we conclude in Sec.~\ref{sec:conclusions}.

\section{Virtual detectors}
\label{sec:virtual_detectors}

\subsection{Fundamentals}

Ionization from a binding potential means that an electron moves away
from the vicinity of the binding potential's minimum.  A virtual
detector allows us to quantify this dynamics.  More specifically, a
virtual detector is a hypothetical device that determines how the
probability to find a particle within some specific space region changes
over time.  In the following, we present its mathematical foundations.

The quantum mechanical evolution of an electron's wave function
$\Psi(\vec r, t)$ with mass $m$ and charge $q$ is governed by the
Schr{\"o}dinger equation
\begin{equation}
  \I\hbar\frac{\partial\Psi(\vec r, t)}{\partial t} =
  \left(
    \frac{1}{2m}\left(
      -\I\hbar\vec\nabla - q \vec A(\vec r, t)
    \right)^2 + q \phi(\vec r, t)
  \right)\Psi(\vec r, t) \,,
\end{equation}
where $\phi(\vec r, t)$ and $\vec A(\vec r, t)$ denote the
electromagnetic potentials.  The probability density to find the
electron at position $\vec r$ at time $t$ is given by
\begin{equation}
  \varrho(\vec r, t) = \Psi(\vec r, t)^*\Psi(\vec r, t) \,,
\end{equation}
where $\Psi(\vec r, t)^*$ indicates the complex conjugate of $\Psi(\vec
r, t)$.  The dynamics of the density $\varrho(\vec r, t)$ is associated
with the probability current 
\begin{equation}
  \vec j(\vec r, t) =
  \frac{1}{m}\Re\big(
    \Psi(\vec r, t)^*\left(
      -\I\hbar\vec\nabla - q \vec A(\vec r, t)
    \right)\Psi(\vec r, t)) 
  \big)\,.
\end{equation}
Expressing the wave function $\Psi(\vec r, t)$ as 
\begin{equation}
  \Psi(\vec r, t) = \sqrt{\varrho(\vec r, t)} \exp(\I\varphi(\vec r, t))\,,
\end{equation}
the probability current $\vec j(\vec r, t)$ may be written as a product
of the probability density $\varrho(\vec r, t)$ and some local velocity,
viz.,
\begin{equation}
  \label{eq:j_rho_v}
  \vec j(\vec r, t) =
  \frac{\varrho(\vec r, t)}{m}\left(
    \hbar\vec\nabla\varphi(\vec r, t) - q \vec A(\vec r, t)
  \right)\,,
\end{equation}
which also occurs in the framework of Bohmian mechanics
\cite{Pladevall:Mompart:2012:Applied_Bohmian}.  The probability density
and the probability current fulfill the continuity equation
\begin{equation}
  \label{eq:continuity}
  - \frac{\partial \varrho(\vec r, t)}{\partial t} = 
  \vec\nabla \cdot \vec j(\vec r, t) \,.
\end{equation}
For any compact subspace $V$ of the physical space with a piecewise smooth
boundary $S$  the continuity equation \eqref{eq:continuity} yields by
integration over the subspace
\begin{equation}
  \label{eq:continuity_int}
  - \frac{\D}{\D t}
  \int_V \varrho(\vec r, t)\,\D V =
  \int_S \vec j(\vec r, t)\cdot\vec n\,\D S \,,
\end{equation}
where $\vec n$ denotes the outward pointing unit normal field of the
boundary $S$.  Equation~\eqref{eq:continuity_int} holds also for
unbounded subspaces if $\varrho(\vec r, t)$ decreases fast enough when
$|\vec r|\to\infty$.  Thus, the rate of the change of the probability to
find the electron within the space region $V$ is given by the
probability current through the volume's surface
\begin{equation}
  D_S(t) = \int_S\vec j(\vec r, t)\cdot\vec n\,\D S\,.
\end{equation}

A virtual detector is a hypothetical device that measures $D_S(t)$.  In
other words, a virtual detector placed at the surface $S$ determines how
much probability passes from one side of the surface to the other per
unit time.  Placing $S$ at sufficient distance around a bound wave
packet, which is ionized by a time-dependent electric field $\vec E(t)$
with $\lim_{t\to\pm\infty}\vec E(t)=0$, then
$\int_{-\infty}^\infty D_S(t)\,\D t$ equals the total ionization
probability.  Furthermore, if $D_S(t)$ is strictly nonnegative or
nonpositive $|D_S(t)|\Delta t$ can be interpreted as the probability
that the electron moves from one side of the surface $S$ to the other
within the short time interval $[t, t+\Delta t]$.  For one-dimensional
systems the surface $S$ degenerates to a single point.  For this case,
it can be shown within the framework of Bohmian mechanics that
$|D_S(t)|$ is up to normalization the probability distribution of the
time of arrival at the point $S$
\cite{Leavens:1993:Arrival_time,McKinnon:Leavens:1995:Distributions_of}.
In general, $|D_S(t)|$ is proportional to the probability that the
electron arrives at the surface $S$ at time $t$.

\begin{figure*}
  \hspace{\fill}%
  \includegraphics[scale=0.5]{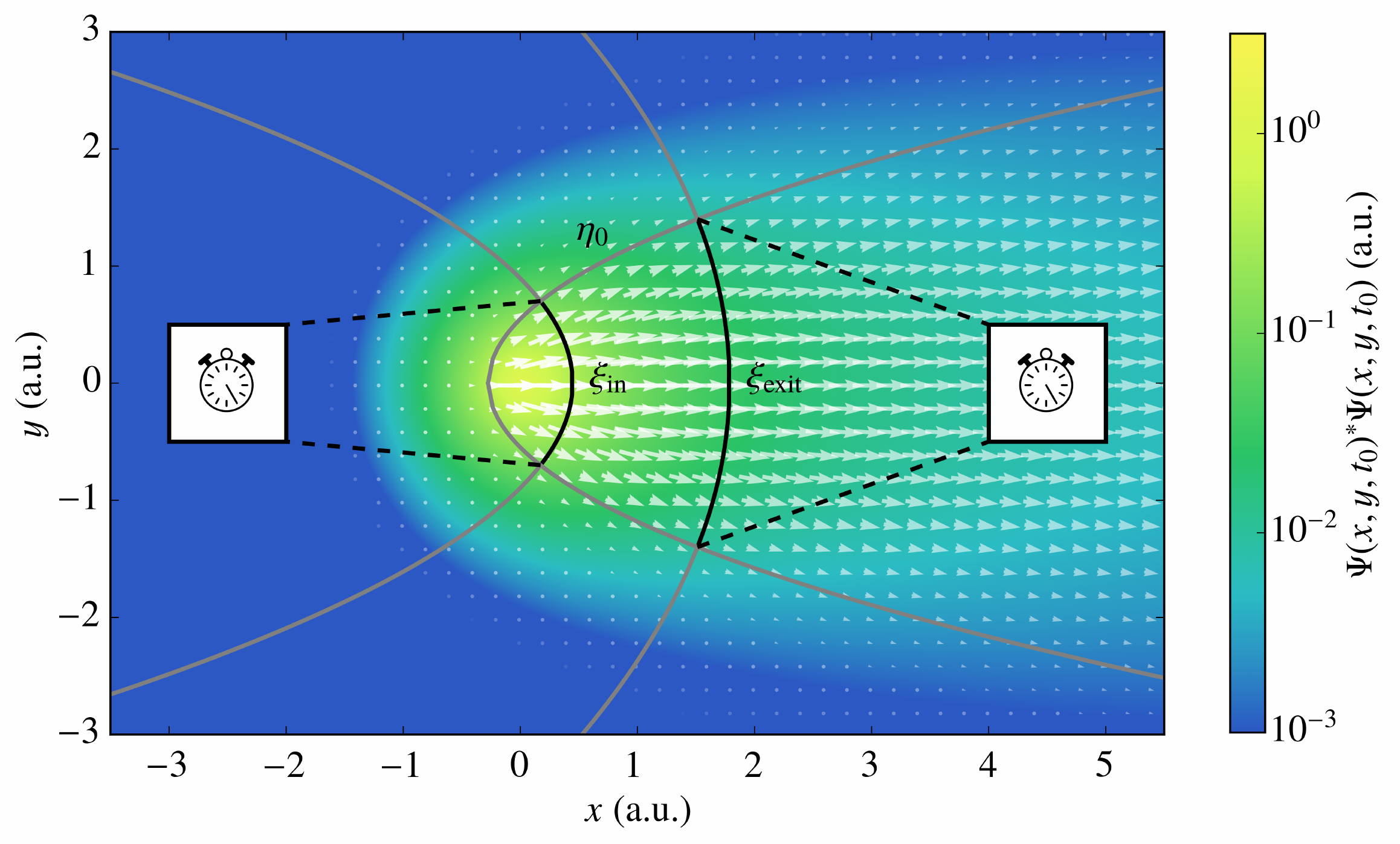}%
  \hspace{\fill}%
  \hspace{\fill}%
  \raisebox{2.2cm}{%
    \includegraphics[scale=0.45]{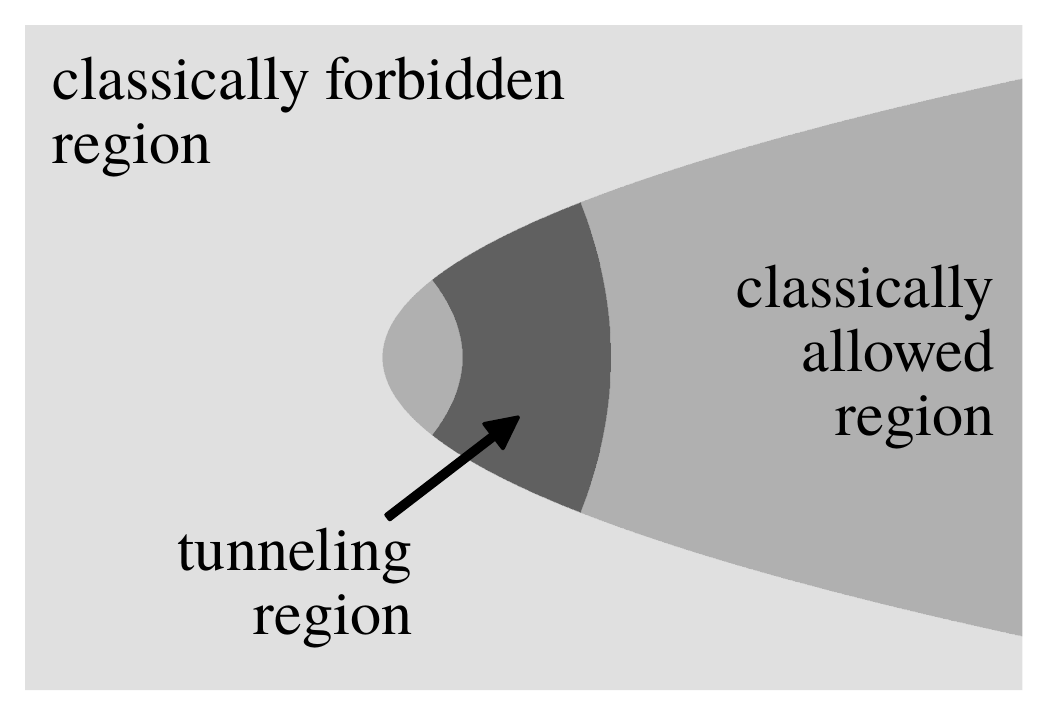}%
  }%
  \hspace{\fill}%
  \mbox{}
  \caption{Schematic of tunneling time determination via virtual
    detectors at the tunneling entry and the tunneling exit.  The lines
    $\xi=\xi_\mathrm{in,out}$ and $\eta=\eta_0$ which determine
    tunneling region, the classically forbidden, and the classical
    allowed regions are indicated by the solid lines; see
    Ref.~\cite{Lantoine:Russell:2011:Complete_closed-form} and the
    \hyperref[sec:appendix]{Appendix} for details.  The black solid
    lines indicate the positions of the virtual detectors.  The
    wave-function's probability density $\Psi(x, y, t)^*\Psi(x, y, t)$
    at the instant of electric field maximum, i.\,e., $t=t_0$, is
    represented by colors for $\mathcal{E}_0=1.1\,\mathrm{a.u.}$ and for
    the Keldysh parameter $\gamma=0.25$.  Arrows indicate the
    probability current $\vec j(x, y, t)$, where the arrows' length is
    proportional to the absolute value of the current and the arrows'
    opacity scales with the wave function's probability density.}
  \label{fig:tunneling_potentials_2d}
\end{figure*}

\subsection{Tunnel ionization from a two-dimensional Coulomb potential}

In the following, we will study tunnel ionization from a two-dimensional
Coulomb potential.  The restriction to two dimensions is mainly because
this system resembles tunnel ionization from hydrogen-like ions while
keeping the computational demands small.  In the long-wavelength limit,
i.\,e., when the dipole approximation is applicable, the
three-dimensional Coulomb potential with an external linearly polarized
electric field has rotational symmetry, which makes this system
effectively two-dimensional.  Furthermore, the two-dimensional Coulomb
problem is of interest in its own and has been used to model some
semiconductor systems
\cite{Ando:Fowler:etal:1982:Electronic_properties,Tanaka:Kobashi:etal:1987:LoSurdo-Stark_effect,Glutsch:Chemla:etal:1996:Numerical_calculation}.
It can also be derived as a limit of the Hamiltonian of a
three-dimensional hydrogen atom in a planar slab as the width of the
slab tends to zero \cite{Duclos:Stovcek:etal:2010:On_two-dimensional}.

Applying the dipole approximation and choosing the coordinate system
such that the linearly polarized external electric field
$\mathcal{E}(t)$ points into the (negative) $x$~direction, the
Schr{\"o}dinger equation reads in atomic units, which are applied for the
remainder of this article,
\begin{multline}
  \label{eq:Schrodinger_2d}
  \I\frac{\partial\Psi(x, y, t)}{\partial t} = \op H \Psi(x, y, t) =\\
  \left(
    -\frac{1}{2}\frac{\partial^2}{\partial x^2}
    -\frac{1}{2}\frac{\partial^2}{\partial y^2}
    -\frac{1}{\sqrt{x^2+y^2}} - x\mathcal{E}(t)
  \right)\Psi(x, y, t)\,.
\end{multline}
In order to study the ionization dynamics without undesirable artifacts,
i.\,e., to avoid multiple ionization and rescattering, we choose an
electric field pulse with a unique maximum,
\begin{equation}
  \label{eq:E_t}
  \mathcal{E}(t)=\mathcal{E}_0\exp\left(
    -\frac{\omega^2(t-t_0)^2}{2}
  \right)\,,
\end{equation}
where $t_0$ denotes the instant of maximal electric field
$\mathcal{E}(t_0)=\mathcal{E}_0$ and
$\tau_{\mathcal{E}} = \sqrt{2}/\omega$ is the time scale of the raise
and decay of the electric field.

Cartesian coordinates are not the most suitable choice to deal with the
Coulomb problem with an external homogeneous electric field.  Parabolic
coordinates are the natural coordinate system for this problem.  In
particular, it allows us to define a tunneling barrier
\cite{Pfeiffer:Cirelli:etal:2011:Attoclock_reveals}.  As one can show,
the Hamiltonian $\hat H$ in Eq.~\eqref{eq:Schrodinger_2d} separates in
parabolic coordinates $0\le\xi<\infty$ and $0\le\eta<\infty$, which are
related to the Cartesian coordinates $x$ and $y$ via
\begin{subequations}
  \label{eq:x_y_xi_eta}
  \begin{align}
    x & = \frac{\xi - \eta}{2} \,, \\
    y & = \sqrt{\xi \eta} \,,
  \end{align}
\end{subequations}
into two independent one-dimensional Schr{\"o}dinger-type Hamiltonians with
specific potentials $V_1(\xi)$ and $V_2(\eta)$.  While $V_2(\eta)$ is
purely attractive, $V_1(\xi)$ represents the tunneling barrier, and,
therefore, $\xi$ denotes the tunneling direction.  Intersections of the
potentials with the energy value $E/4$ of the one-dimensional
Hamiltonians define the borders between the classically allowed, the
tunneling, and the classically forbidden regions.  Here, $E$ denotes the
Stark-effect corrected ground state energy of the two-dimensional
system.  The classically allowed region, the tunneling region, and the
classically forbidden region are characterized by
\begin{subequations}
  \begin{align}
    E/4 & > V_1(\xi)\,, & E/4 & > V_2(\eta)\,, \\
    E/4 & > V_1(\xi)\,, & E/4 & < V_2(\eta)\,, \\
    E/4 & < V_1(\xi)\,, & E/4 & < V_2(\eta)\,,
  \end{align}
\end{subequations}
respectively.  The tunneling barrier is confined by the three parabolas
at $\xi=\xi_\mathrm{in}$, $\xi=\xi_\mathrm{exit}$, and $\eta=\eta_0$,
which are functions of the applied electric field strength
$\mathcal{E}(t)$.  See the \hyperref[sec:appendix]{Appendix} for details
and Figs.~\ref{fig:tunneling_potentials}
and~\ref{fig:tunneling_potentials_2d}.

The virtual detectors are placed at $\xi=\xi_\mathrm{in}$ and
$\xi=\xi_\mathrm{exit}$ as given for the maximal electric field strength
$\mathcal{E}_0$ at $t=t_0$.  In two dimensions, the surface integral in
Eq.~\eqref{eq:continuity_int} becomes a line integral of a vector field
$\vec j(x, y, t)$.  Thus, the probability current over the entry line
$\xi=\xi_\mathrm{in}$, denoted by $D_{\xi_\mathrm{in}}(t)$, and the
current over the exit line $\xi=\xi_\mathrm{exit}$, denoted by
$D_{\xi_\mathrm{exit}}(t)$, are given explicitly by
\begin{multline}
  \label{eq:D_xi}
  D_{\xi}(t) = 
  \int_0^{\eta_0} 
  \vec j(x(\xi, \eta), y(\xi, \eta), t)
  \cdot
  \begin{pmatrix}
    \frac{\partial x(\xi, \eta)}{\partial \eta} \\[1.5ex]
    \frac{\partial y(\xi, \eta)}{\partial \eta} 
  \end{pmatrix}\D\eta\\ + 
  \int_0^{\eta_0} 
  \vec j(x(\xi, \eta), -y(\xi, \eta), t)
  \cdot
  \begin{pmatrix}
    \frac{\partial x(\xi, \eta)}{\partial \eta} \\[1.5ex]
    -\frac{\partial y(\xi, \eta)}{\partial \eta} 
  \end{pmatrix}\D\eta \,,
\end{multline}
where we have taken into account that Eq.~\eqref{eq:x_y_xi_eta} covers
only the upper half of the coordinate system.  Monitoring the
probability current at these \emph{fixed} entry and exit lines can be
justified as follows.  For static fields the tunneling probability is
maximal for maximal electric fields, and it is exponentially suppressed
for lower fields.  Furthermore, the applied electric field
\eqref{eq:E_t} is quasistatic for $|t-t_0|<\tau_\mathcal{E}$:
\begin{equation}
  \mathcal{E}(t) = 
  \mathcal{E}_0\left(
    1- (t-t_0)^2/\tau_\mathcal{E}^2 + 
    O\left((t-t_0)^4/\tau_\mathcal{E}^4\right)
  \right)\,.
\end{equation}
Therefore, the tunneling barrier is also quasistatic if times close to
$t_0$ are considered.  As it will be shown later, the extracted
tunneling times are short compared to $\tau_\mathcal{E}$ indeed.

\section{Numerical results and discussion}
\label{sec:results}

\subsection{Entry time, exit time, and tunneling time}

The Schr{\"o}dinger equation~\eqref{eq:Schrodinger_2d} is solved numerically
by employing a Lanczos propagator
\cite{Park:Light:1986:Unitary_quantum,Beerwerth:Bauke:2015:Krylov_subspace}
and fourth-order finite differences for the discretization of the
Schr{\"o}dinger Hamiltonian.  The ground state of the two-dimensional
Coulomb potential was chosen as an initial condition at
$t-t_0=-5\tau_{\mathcal{E}}$, i.\,e., when the external electric field
is negligible small.  The so-called Keldysh parameter
$\gamma=\omega\sqrt{-2E_0}/\mathcal{E}_0$ \cite{Keldysh:1965:Ionization}
characterizes the ionization process as dominated by tunneling for
$\gamma\ll1$ and by multiphoton ionization for $\gamma\gg1$. Here, $E_0$
denotes the ground state binding energy, which equals
$E_0=-2\,\mathrm{a.u.}$ for the two-dimensional Coulomb problem; see
also the \hyperref[sec:appendix]{Appendix}.  In the following, the
electric field amplitude $\mathcal{E}_0$ and the frequency $\omega$ are
adjusted such that $\gamma=0.25<1$.

Figure~\ref{fig:tunneling_potentials_2d} illustrates the ionization
dynamics in the vicinity of the tunneling barrier by presenting the
electron's probability density $\Psi(x, y, t)^*\Psi(x, y, t)$ and the
probability current $\vec j(x, y, t)$ at the instant of maximal field
strength $t=t_0$.  The external electric field induces a probability
current opposite to the direction of the electric field.  This flow is
nonnegligible only in the region $\eta<\eta_0$, i.\,e., in the
classically allowed region and the tunneling region.  Thus, the shape of
the probability current confirms that parabolic coordinates are the
natural choice to define a tunneling barrier for the Coulomb problem.

\begin{figure}
  \centering
  \includegraphics[scale=0.45]{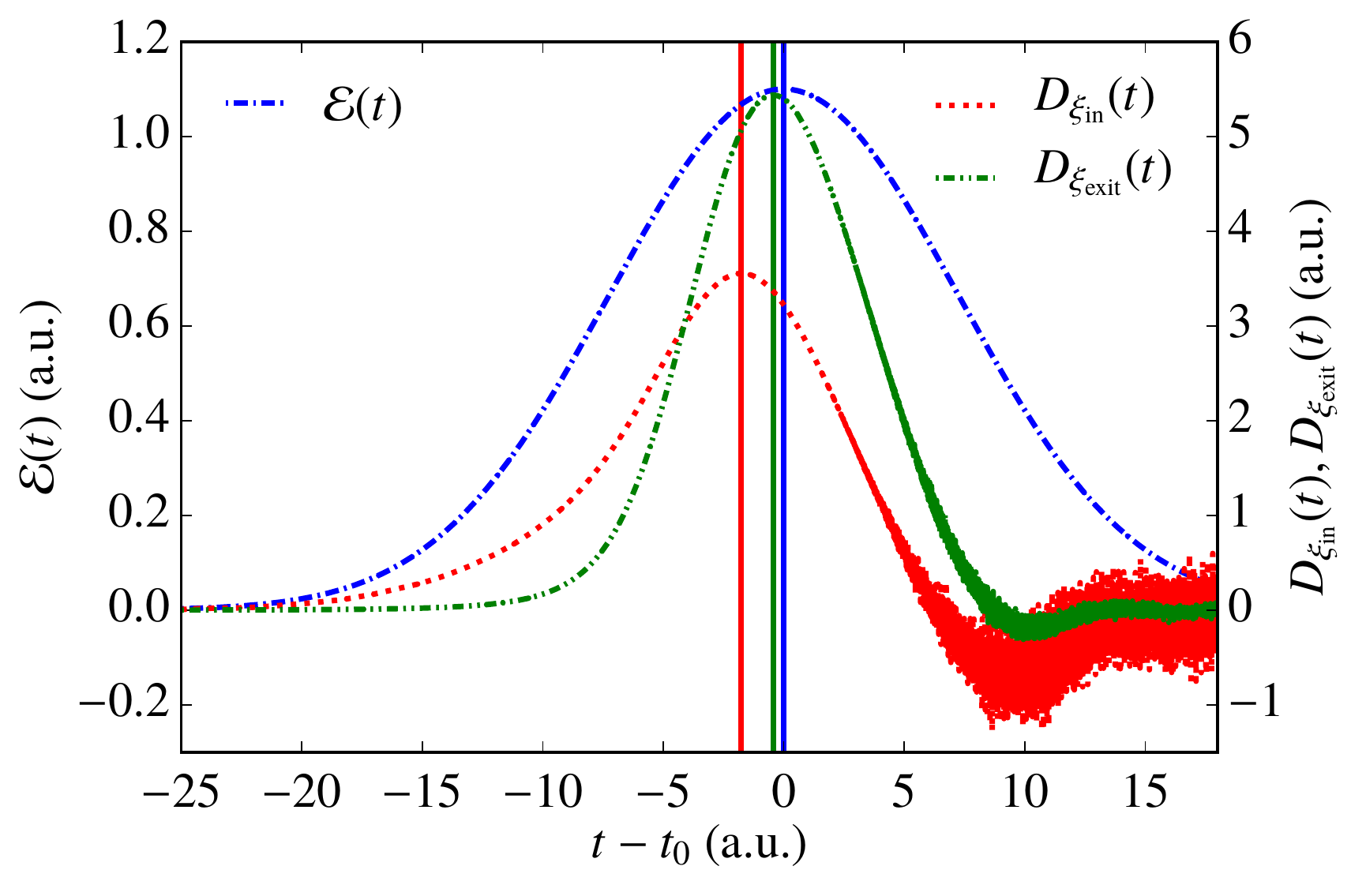}
  \caption{The probability current $D_{\xi_\mathrm{in}}(t)$ over the
    entry line $\xi=\xi_\mathrm{in}$ and the current over
    $D_{\xi_\mathrm{exit}}(t)$ the exit line $\xi=\xi_\mathrm{exit}$ and
    the amplitude of the external electric field $\mathcal{E}(t)$ as
    functions of time $t$.  The three vertical lines indicate from left
    to right the moments of the maximal probability current over the
    line $\xi=\xi_\mathrm{in}$, of the maximal electric field, and of
    maximal probability current over the line $\xi=\xi_\mathrm{exit}$.
    The parameters of the electric field correspond to a maximal field
    strengths of $\mathcal{E}_0=1.1\,\mathrm{a.u.}$ and a Keldysh
    parameter $\gamma=0.25$.}
  \label{fig:flow}
\end{figure}

The action of the external electric field induces a probability current,
which flows over the tunneling barrier's entry line
$\xi=\xi_\mathrm{in}$ and its exit line $\xi=\xi_\mathrm{exit}$.  This
means that probability is carried over from the center region of the
Coulomb potential into the tunneling barrier, from where it can escape
into the classically allowed region.  The quantities
$D_{\xi_\mathrm{in}}(t)$ and $D_{\xi_\mathrm{exit}}(t)$ grow and decay
over time with the applied external electric field, as shown in
Fig.~\ref{fig:flow}.  As a central result of our numerical solution of
the Schr{\"o}dinger equation, however, the position of the maxima of
$D_{\xi_\mathrm{in}}(t)$, $D_{\xi_\mathrm{exit}}(t)$, and the electric
field $\mathcal{E}(t)$ do not coincide.  The probability current over
the tunneling entry $D_{\xi_\mathrm{in}}(t)$ reaches its maximum before
the maximum of the electric field is attained.  Furthermore, the
probability current over the tunneling exit $D_{\xi_\mathrm{exit}}(t)$
may reach its maximum before or after the maximum of the electric field
is attained depending on the electric field strength $\mathcal{E}_0$.
For the rather strong electric field of the parameters of
Fig.~\ref{fig:flow}, $D_{\xi_\mathrm{exit}}(t)$ reaches its maximum
slightly before the electric field.  Turning off the external electric
field, the flow at the tunneling entry becomes negative and then
oscillates rapidly around zero after switching off the electric field.
These oscillations are a result of the excitation of the the bound
portion of the wave function during the action of the electric field.
The final bound state is a superposition of several eigenstates causing
a nonsteady probability current.  In contrast, the quantity
$D_{\xi_\mathrm{exit}}(t)$ remains positive as reflections are absent at
the tunneling exit and behind the tunneling barrier
\cite{Teeny:etal:2016:Ionization_Time}.

Monitoring the probability current at $\xi=\xi_\mathrm{in}$ and
$\xi=\xi_\mathrm{exit}$ allows us to determine the moments when the
electrons enters or leaves the tunneling barrier with maximal
probability.  The probability to cross the tunneling barrier entry at
$\xi=\xi_\mathrm{in}$ or the exit at $\xi=\xi_\mathrm{exit}$ during the
small time interval $[t, t+\Delta t]$ is
$\Delta t\,D_{\xi_\mathrm{in}}(t)$ and
$\Delta t\,D_{\xi_\mathrm{exit}}(t)$, respectively.  This motivates us
to introduce the times of maximal flow over the tunneling entry and
exit, i.\,e., the positions of the maxima of the arrival-time
distributions at the tunneling entry and exit:
\begin{align}
  \label{eq:t_in}
  t_\mathrm{in} & = \argmax D_{\xi_\mathrm{in}}(t) \\
  \intertext{and}
  \label{eq:t_exit}
  t_\mathrm{exit} & = \argmax D_{\xi_\mathrm{exit}}(t) \,.
\end{align}
Then the delay between the instant of the maximum of the driving
electric field $t_0$ and the exit time $t_\mathrm{exit}$, the tunneling
delay, is
\begin{align}
  \tau_\mathrm{exit} & = t_\mathrm{exit} - t_0 \,, \\
  \intertext{and the tunneling time}
  \tau_\mathrm{tsub} & = t_\mathrm{exit} - t_\mathrm{in}
\end{align}
can be introduced with the definitions above.  A measurement of the
delay $\tau_\mathrm{exit}$ is implemented in the attoclock experiments.
The time $\tau_\mathrm{tsub}$ denotes the delay between the instants
when the probability flow is maximal at the tunneling barrier's entry
and at the exit.  Thus, $\tau_\mathrm{tsub}$ may be interpreted as the
typical time which an electron needs to pass the tunneling barrier.

\begin{figure}
  \centering
  \includegraphics[scale=0.45]{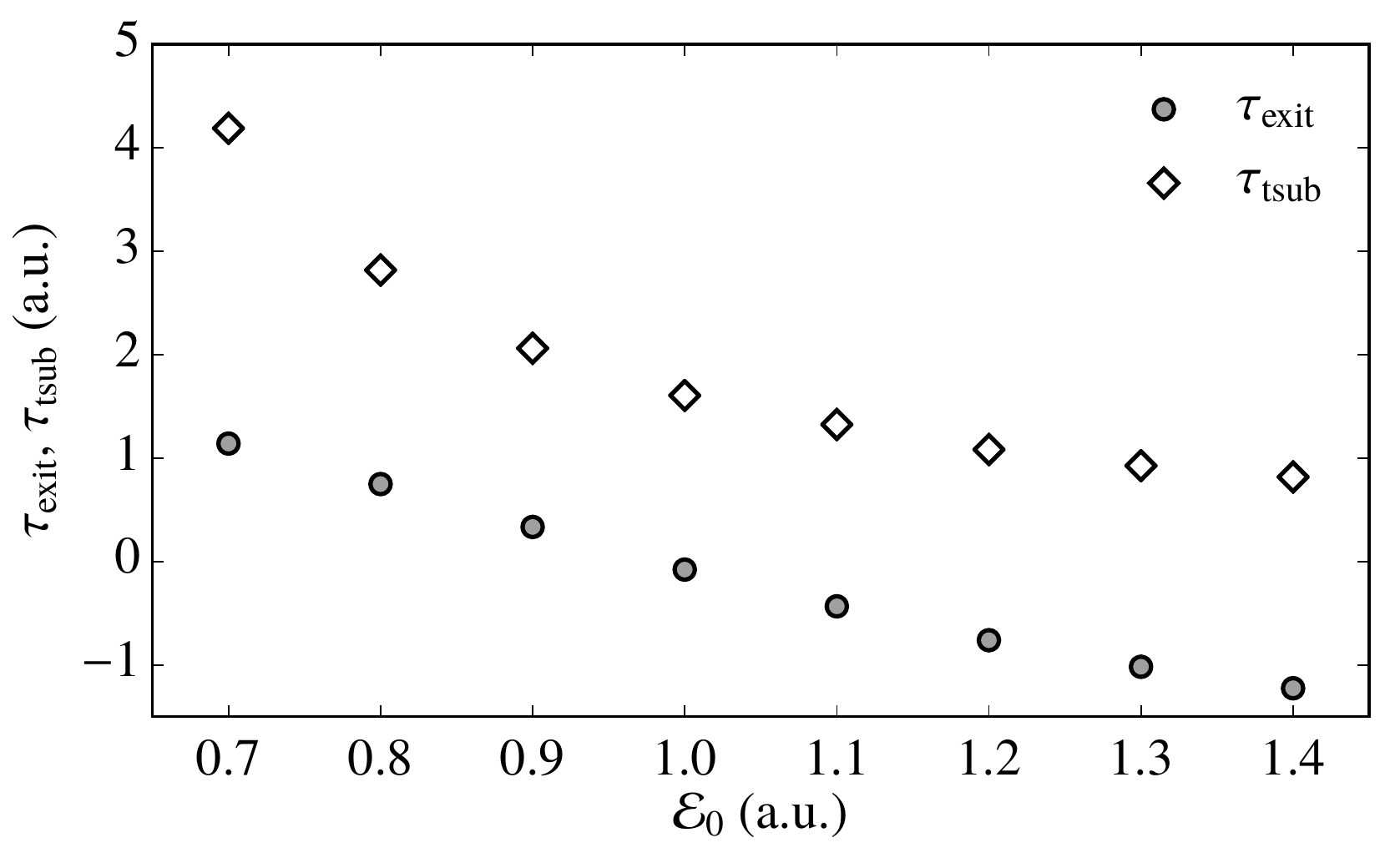}
  \caption{The tunneling delay $\tau_\mathrm{tsub}$ and the tunneling
    time $\tau_\mathrm{exit}$ as functions of the electric field
    amplitude $\mathcal{E}_0$.  The frequency $\omega$ is adjusted such
    that the Keldysh parameter is fixed to $\gamma=0.25$.}
  \label{fig:tsub-texit}
\end{figure}

The times $\tau_\mathrm{exit}$ and $\tau_\mathrm{tsub}$ are presented in
Fig.~\ref{fig:tsub-texit} for the considered setup and varying electric
field amplitudes $\mathcal{E}_0$.  An increasing electric field
amplitude reduces the width of the tunneling barrier.  Consequently, the
tunneling time $\tau_\mathrm{tsub}$ decreases with increasing electric
field amplitude $\mathcal{E}_0$, as shown in Fig.~\ref{fig:tsub-texit}.
The time $\tau_\mathrm{tsub}$ remains, however, always positive.  This
means that the probability current reaches its maximum at the entry of
the tunneling barrier before the current becomes maximal at the exit,
i.\,e., the electron enters the barrier before it exits the barrier.
Furthermore, it follows from the data in Fig.~\ref{fig:tsub-texit} that
$\tau_\mathrm{tsub}>\tau_\mathrm{exit}$ and consequently
$t_\mathrm{entry}<t_0$, because
$\tau_\mathrm{tsub}-\tau_\mathrm{exit}=t_0-t_\mathrm{entry}$.  This
means that the electron always enters the tunneling barrier before the
electric field reaches its maximum.

\begin{figure}
  \centering
  \includegraphics[scale=0.45]{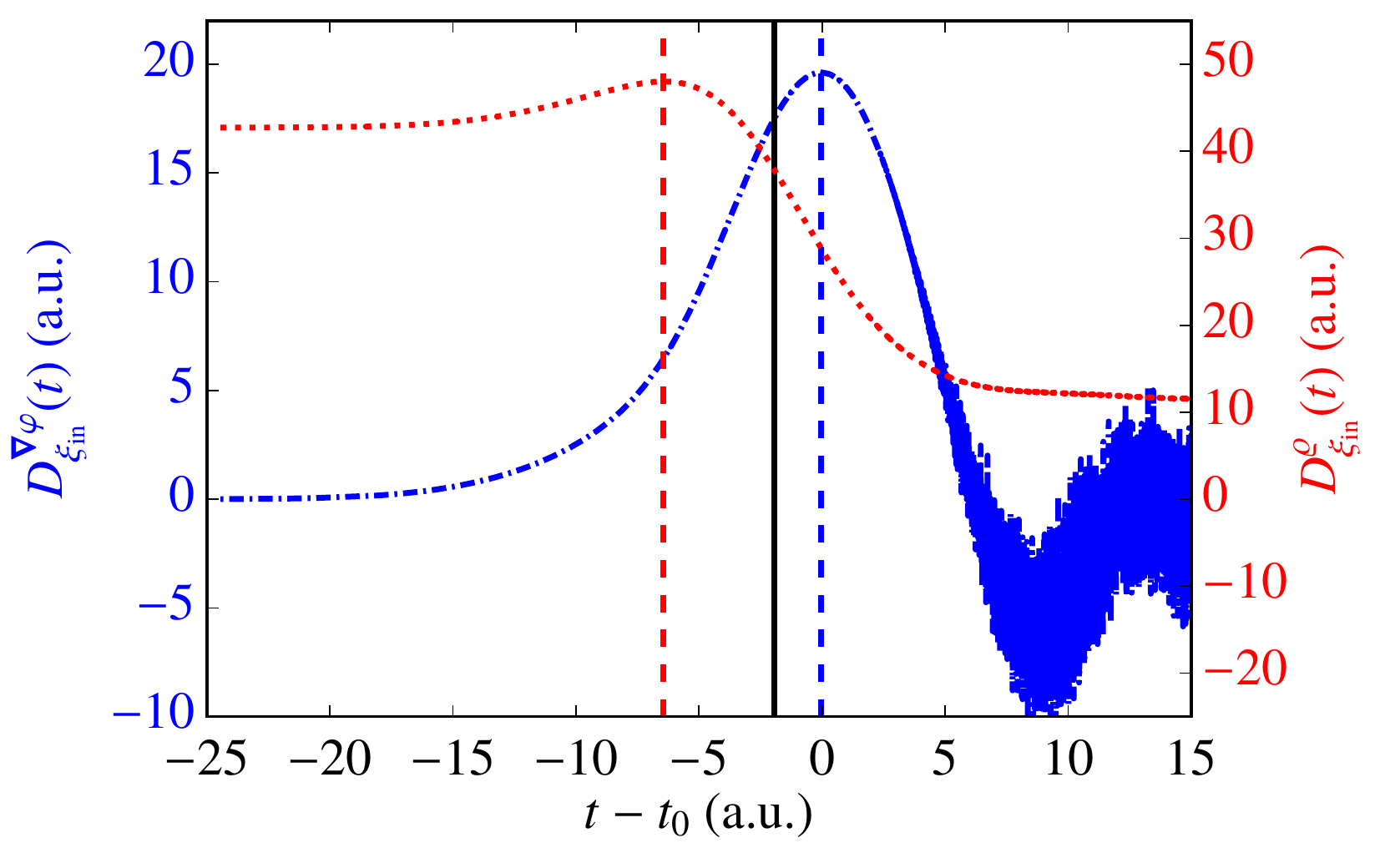}
  \caption{The probability density $D_{\xi_\mathrm{in}}^{\varrho}(t)$
    (red dashed line, right scale) and the local velocity
    $D_{\xi_\mathrm{in}}^{\vec\nabla\varphi}(t)$ (blue dot-dashed line,
    left scale) along the entry line $\xi=\xi_\mathrm{in}$ are plotted
    as a function of time for the electric field strength
    $\mathcal{E}_0=1.3\,\mathrm{a.u.}$ The solid black line denotes the
    instant of maximal $D_{\xi_\mathrm{in}}(t)$, vertical dashed lines
    indicate the position of the maxima of
    $D_{\xi_\mathrm{in}}^{\varrho}(t)$ and
    $D_{\xi_\mathrm{in}}^{\vec\nabla\varphi}(t)$.}
  \label{fig:fvr}
\end{figure}

To explain the reason for this behavior, we define in analogy to
Eq.~\eqref{eq:D_xi} the two integrals over the probability density
\begin{multline}
  D_{\xi}^{\varrho}(t) = 
  \int_0^{\eta_0} 
  \varrho(x(\xi, \eta), y(\xi, \eta), t) \,\D\eta \\ + 
  \int_0^{\eta_0}
  \varrho(x(\xi, \eta), -y(\xi, \eta), t) \,\D\eta
\end{multline}
and over the velocity field\pagebreak[1]
\begin{multline}
  D_{\xi}^{\vec\nabla\varphi}(t) = 
  \int_0^{\eta_0} 
  \vec \nabla\varphi(x(\xi, \eta), y(\xi, \eta), t)
  \cdot
  \begin{pmatrix}
    \frac{\partial x(\xi, \eta)}{\partial \eta} \\[1.5ex]
    \frac{\partial y(\xi, \eta)}{\partial \eta} 
  \end{pmatrix}\,\D\eta \\ +
  \int_0^{\eta_0} 
  \vec \nabla\varphi(x(\xi, \eta), -y(\xi, \eta), t)
  \cdot
  \begin{pmatrix}
    \frac{\partial x(\xi, \eta)}{\partial \eta} \\[1.5ex]
    -\frac{\partial y(\xi, \eta)}{\partial \eta} 
  \end{pmatrix}\,\D\eta\,,
\end{multline}
which are motivated by fact that the probability current can be written
as a product of the probability density and the local velocity; see
Eq.~\eqref{eq:j_rho_v}.  $D_{\xi_\mathrm{in}}^{\varrho}(t)$ gives the
probability density integrated along the entry line
$\xi=\xi_\mathrm{in}$ as a function of time, while
$D_{\xi_\mathrm{in}}^{\vec\nabla\varphi}(t)$ denotes the integrated
velocity.  As shown in Fig.~\ref{fig:fvr}, the velocity along the entry
line $D_{\xi_\mathrm{in}}^{\vec\nabla\varphi}(t)$ increases with the
electric field, reaches a maximum at the instant of electric field
maximum $t=t_0$, then decreases with the electric field, and it becomes
even negative due to reflections from the tunneling process.  Finally,
the integrated velocity $D_{\xi_\mathrm{in}}^{\vec\nabla\varphi}(t)$
oscillates around zero indicating the excitation of the bound state.
The probability density along the entry line $D_{\xi_{in}}^{\varrho}(t)$
is nonzero even at $t\ll t_0$ because the wave function penetrates the
tunneling barrier even for $\mathcal{E}=0$.  Increasing the electric
field, the probability density flow from the core into the barrier
increases the probability density along the entry line.  Although the
local velocity at the tunneling entry is maximal at $t=t_0$ in sync with
the electric field, the probability flow along the entry line
$D_{\xi_\mathrm{in}}^{\varrho}(t)$ reaches its maximum earlier.  The
quantity $D_{\xi_\mathrm{in}}^{\varrho}(t)$ goes asymptotically to a
constant value less than the initial value because some of the total
probability has tunneled.  Because the probability flow is the product
of the probability density and the local velocity, the maximum of the
integrated probability flow $D_{\xi_\mathrm{in}}(t)$ is reached between
the instants of the maxima of the integrated density
$D_{\xi_\mathrm{in}}^{\varrho}(t)$ and the integrated velocity
$D_{\xi_\mathrm{in}}^{\vec\nabla\varphi}(t)$; this means
$t_\mathrm{in}<t_0$.  The time $t_\mathrm{exit}$ is usually larger than
$t_0$, i.\,e., $\tau_\mathrm{exit}>0$, especially for weak external
fields.

After the electron has passed the entry line of the tunneling barrier,
it needs some time to cross the barrier and to reach the tunneling exit.
This means, $t_\mathrm{exit}>t_\mathrm{in}$, but not necessarily
$t_\mathrm{exit}>t_0$.  The latter may be understood by realizing that
the above explanation for $t_\mathrm{in}<t_0$ is not specific to the
coordinate $\xi=\xi_\mathrm{in}$.  In particular, it also applies to
$\xi=\xi_\mathrm{exit}$.  Consequently, also the integrated flow at the
exit coordinate $D_{\xi_\mathrm{exit}}^{\varrho}(t)$ may reach its
maximum before the maximum of the electric field, as, for example, for
the parameters in Fig.~\ref{fig:flow}.  In general, however, we find
that $D_{\xi_\mathrm{exit}}^{\vec\nabla\varphi}(t)$ is maximal at an
instant very close to $t_0$.

Taking the results above into consideration, the time $t_\mathrm{in}$
should be regarded as a reference for the tunneling time, but not $t_0$.
The latter leads to the delay $\tau_\mathrm{exit}$.  Choosing $t_0$ as a
reference can lead to negative delays $\tau_\mathrm{exit}$ for large
electric field strengths, e.\,g., for
$\mathcal{E}_0\ge1.0\,\mathrm{a.u.}$ for the setup that was applied for
Fig.~\ref{fig:tsub-texit}.  This, however, should not be misinterpreted
as a negative tunneling time because $\tau_\mathrm{exit}$ is not related
to the time which an electron needs to pass the tunneling barrier.  This
time is given by $\tau_\mathrm{tsub}$.  As $\tau_\mathrm{tsub}$ is
always larger than $\tau_\mathrm{exit}$, the delay $\tau_\mathrm{exit}$
may be used as a lower bound of the tunneling time.

\begin{figure}
  \centering
  \includegraphics[scale=0.45]{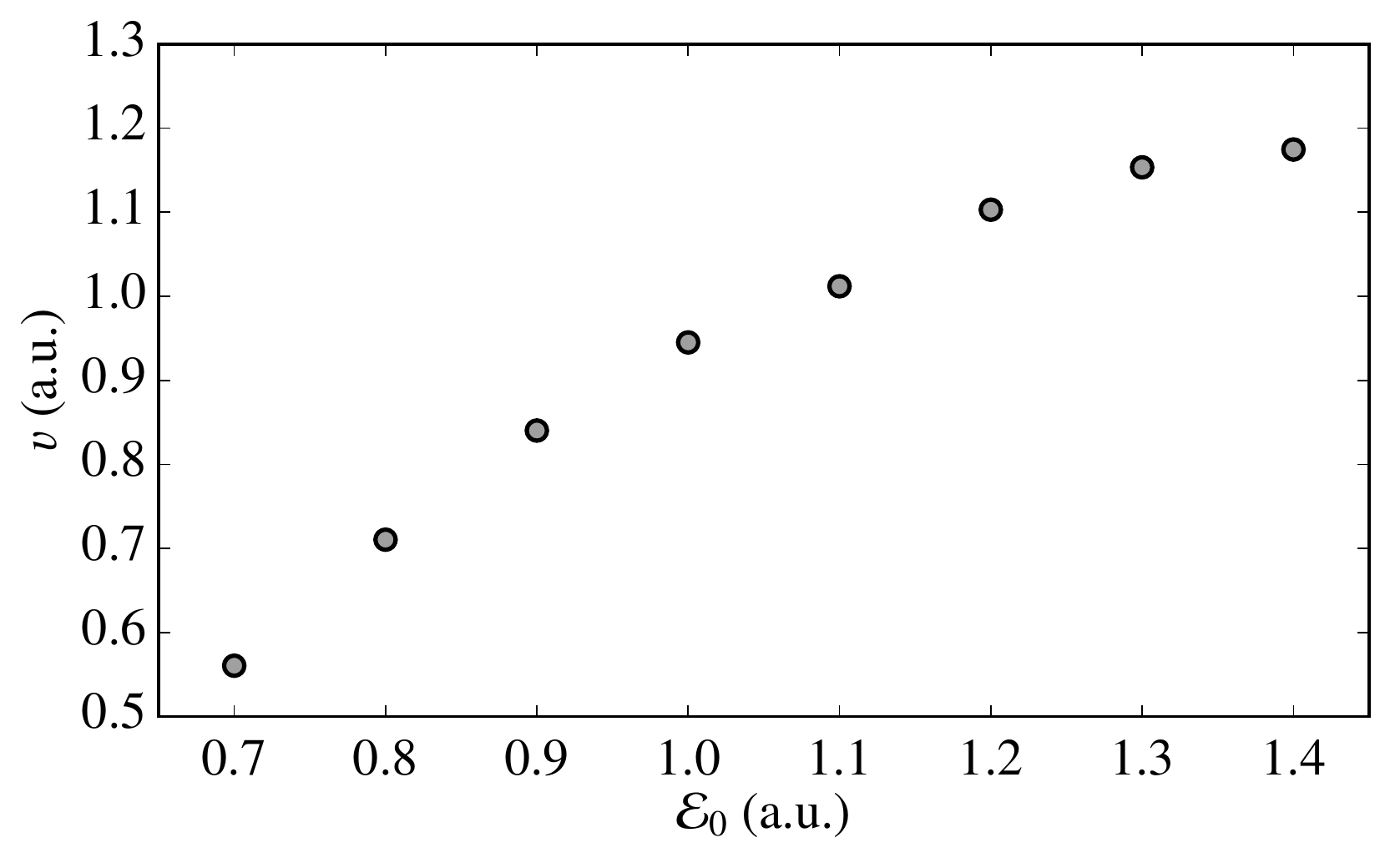}
  \caption{Average velocity $v$ of the motion of the
    probability current's maximum from the tunneling barrier entry to
    the exit as a function of the applied maximal electric field
    strength $\mathcal{E}_0$.}
  \label{fig:electron-speed}
\end{figure}

It has been speculated that tunnel ionization may be instantaneous
\cite{Eckle:Pfeiffer:etal:2008:Attosecond_Ionization,Torlina:Morales:etal:2015:Interpreting_attoclock},
which would be associated with a superluminal traversal of the tunneling
barrier.  Our results, however, clearly indicate that there is a
nonvanishing tunneling time $\tau_\mathrm{tsub}$.  This leads to a
finite average velocity of the motion of from $\xi=\xi_\mathrm{in}$ to
$\xi=\xi_\mathrm{exit}$:
\begin{equation}
  \label{eq:v}
  v = \frac{\xi_\mathrm{exit}-\xi_\mathrm{in}}{2\tau_\mathrm{tsub}} \,.
\end{equation}
Here we have taken into account that the lines of constant $\xi$ are
parabolas in the \mbox{$x$-$y$} plane and the probability current flows
mainly at $y\approx 0$.  As shown in Fig.~\ref{fig:electron-speed},
there is a monotonous relation between the applied maximal electric
field strength $\mathcal{E}_0$ and the velocity $v$.  The velocity
presented in Fig.~\ref{fig:electron-speed}, however, is more than two
orders of magnitude below the speed of light.  Identifying the tunneling
motion of the electron over the tunneling barrier with the motion of the
probability-current's maximum, it is justified to say that tunneling
ionization neither is instantaneous nor moves the electron at
superluminal speed.

\subsection{Quantum and classical trajectories}

\begin{figure}
  \centering
  \includegraphics[scale=0.45]{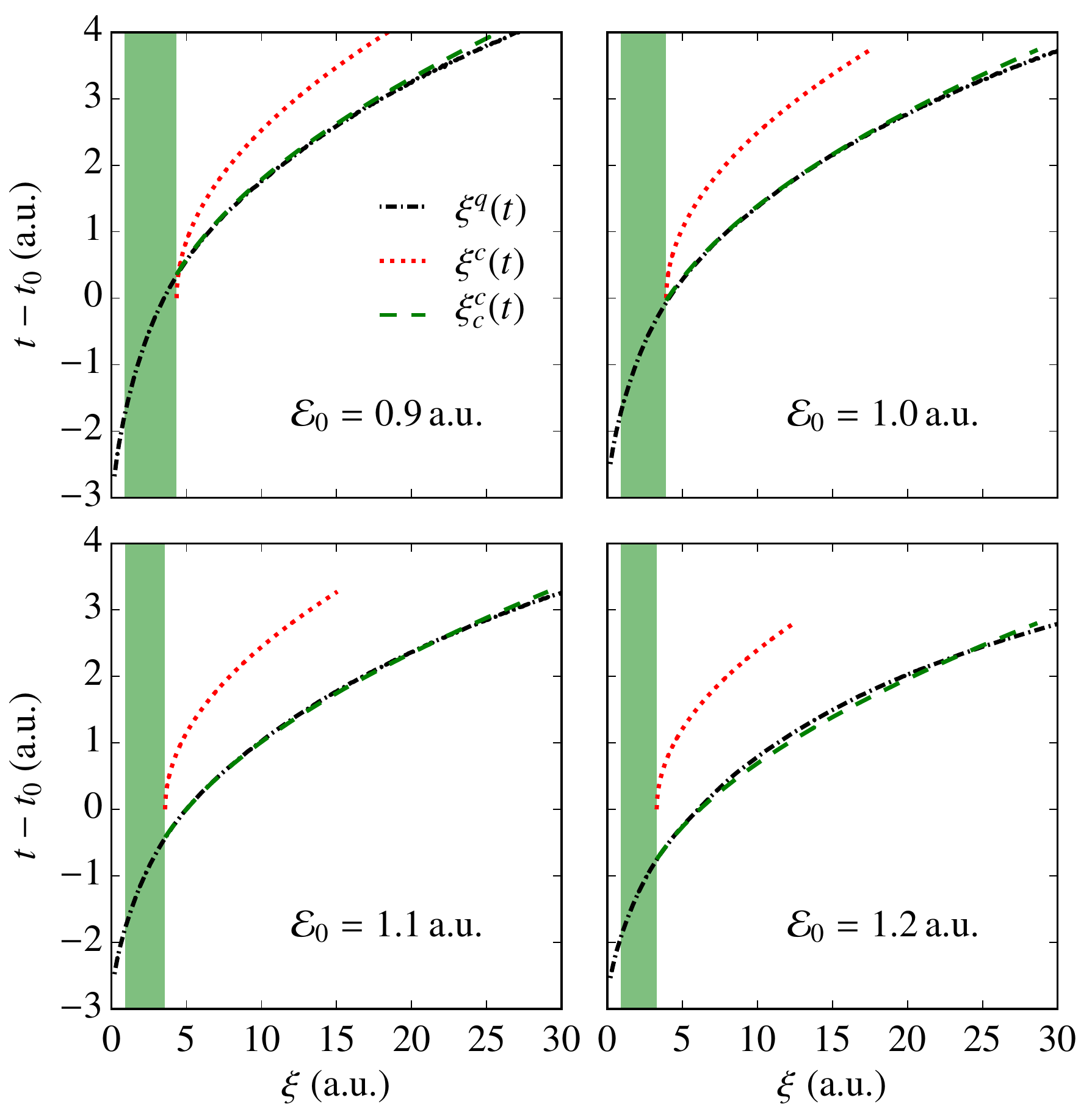}
  \caption{The quantum trajectory of the electron in the $\xi$
    coordinate $\xi^q(t)$, compared to the classical trajectory
    according based on the two-step model $\xi^c(t)$. Both are also
    compared to the classical trajectory corrected according to quantum
    initial conditions $\xi_c^c(t)$. For more information see the main
    text. The shadowed area marks the positions $\xi$ corresponding to
    under the barrier. }
  \label{fig:trajectories}
\end{figure}

The probability currents over different lines of constant $\xi$ reach
their maxima at different times, which allows us to define a quantum
trajectory $\xi^q(t)$ by inverting
\begin{equation}
\label{eq:t_chi}
  t_\xi = \argmax D_\xi(t) \,.
\end{equation}
The quantum trajectory $\xi^q(t)$ defined by Eq.~\eqref{eq:t_chi} can be
compared to the trajectory predicted by a Coulomb-corrected two-step
model $\xi^c(t)$.  This classical trajectory is determined by the Newton
equation
\begin{equation}
  \frac{\D^2\xi^c(t)}{\D t^2} = -\frac{8}{{\xi^c(t)}^2}+2\mathcal{E}(t)\,,
\end{equation}
where the reduction to one dimension is justified as the most probable
trajectory is along the $x$~direction at $y\approx 0$. The initial
conditions for the two-step model are
\begin{subequations}
  \begin{align}
    \xi^c(t_0) & = \xi_\mathrm{exit} \,,\\
    \left.\frac{\D\xi^c(t)}{\D t}\right|_{t=t_0} & = 0 \,,
  \end{align}
\end{subequations} 
meaning that the electron exits at the instant of electric field maximum
$t=t_0$ with zero initial momentum at the turning point
$x=\xi_\mathrm{exit}/2$.  As shown in Fig.~\ref{fig:trajectories} for
different electric field strengths, the classical trajectory $\xi^c(t)$
deviates from the quantum trajectory $\xi^q(t)$ not only near the
barrier exit but also at a far away detector, i.\,e., for
$\xi\gg\xi_\mathrm{exit}$.

\begin{figure}
  \centering
  \includegraphics[scale=0.45]{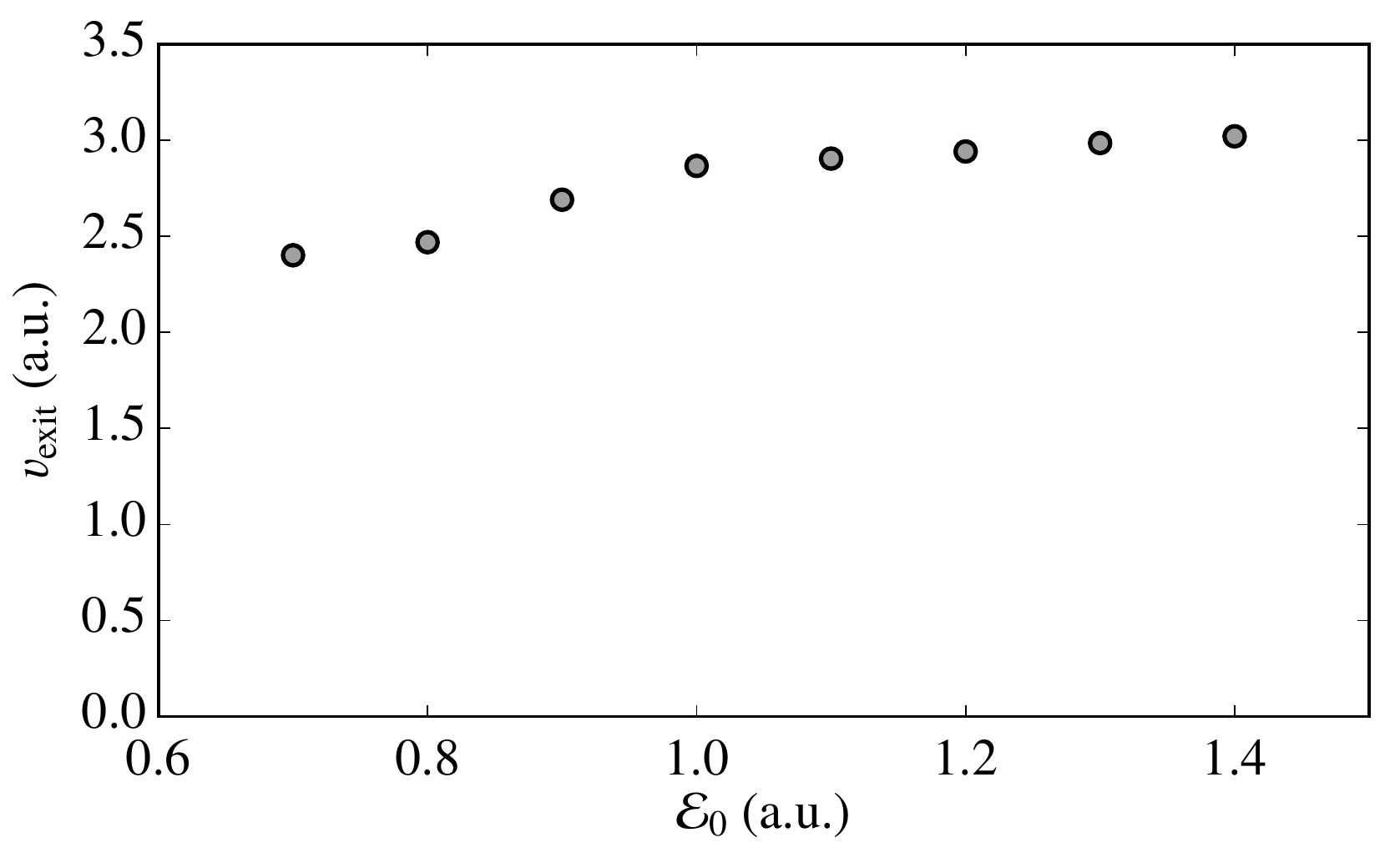}
  \caption{The quantum trajectory's initial velocity $v_\mathrm{exit}$
    plotted for different electric field strengths $\mathcal{E}_0$ for
    Keldysh parameter $\gamma=0.25$.}
  \label{fig:initial}
\end{figure}

From the above studies we have seen that the electron exits the
tunneling barrier at $t_\mathrm{exit}$, which differs from $t_0$.  It
has been shown by quantum mechanical calculations
\cite{Teeny:etal:2016:Ionization_Time} that the electron exits also with
an initial momentum in the electric field direction.  This initial
velocity can be inferred from the quantum trajectory, via
\begin{equation}
  v_\mathrm{exit} = \left.\frac{\D\xi^q(t)}{\D t}\right|_{t=t_\mathrm{exit}} \,,
\end{equation} 
which yields results that are consistent with the results of
Ref.~\cite{Teeny:etal:2016:Ionization_Time}.  The classical trajectory
$\xi^c(t)$ can be corrected by the quantum initial conditions:
\begin{subequations}
  \begin{align}
    \xi^c_c(t_\mathrm{exit}) & = \xi_\mathrm{exit} \,, \\
    \left.\frac{\D\xi^c_c(t)}{\D t}\right|_{t=t_\mathrm{exit}} & = v_\mathrm{exit} \,.
  \end{align}
\end{subequations} 
As shown in Fig.~\ref{fig:trajectories} the corrected classical
trajectory $\xi^c_c(t)$ agrees well with the quantum trajectory
$\xi^q(t)$.  Nevertheless, there is still a slight discrepancy between
both trajectories at a far away detector, which can be attributed to
that $\xi^q(t)$ originates from the motion of a broad wave packet which
experiences a spatially varying Coulomb force
\cite{Ehrenfest:1927:Bemerkung_uber}.  The initial velocity
$v_\mathrm{exit}$ is plotted in Fig.~\ref{fig:initial} for different
electric field strengths.  In agreement with
Ref.~\cite{Teeny:etal:2016:Ionization_Time}, the initial momentum is
slightly dependent on the electric field strength also in the considered
two-dimensional system.  The initial momentum is of the order of the
width of the ground state distribution in momentum space.

The initial velocity $v_\mathrm{exit}$ is difficult to measure directly
in an experiment.  But it also affects the asymptotic velocity of the
ionized electron.  Note that one can choose an electric field strength
such that the quantum trajectory is at $\xi^q=\xi_\mathrm{exit}$ at
$t=t_0$, i.\,e., the electron leaves the tunneling barrier at the moment
of electric-field maximum, which is one assumption of the two-step
model.  In this case, the difference between the electron's actual
asymptotic momentum and the prediction of the classical
Coulomb-corrected two-step model is just the electron's initial momentum
at the barrier exit. For the two-dimensional model as considered in
Fig.~\ref{fig:trajectories} the condition $\xi^q(t_0)=\xi_\mathrm{exit}$
is fulfilled for an electric field strength $\mathcal{E}_0=1.0$\,a.u.

\section{Conclusions}
\label{sec:conclusions}

We studied the dynamics of tunnel ionization by analyzing the quantum
mechanical wave-function's probability current via virtual detectors.
Virtual detectors are a capable concept for determining tunneling times
because they allow us to identify well-defined moments when the ionized
electron enters and leaves the tunneling barrier.

The numerical solutions of the time-dependent Schr{\"o}dinger equation and
the dynamics of the probability current at the tunneling barrier
indicate that the electron spends a nonvanishing time under the
potential barrier.  This corresponds to a crossing velocity that is much
lower than the speed of light.  Furthermore, the electron may enter the
classical forbidden region on average before the instant of the maximum
of the driving electric field.  Therefore, the instant of electric field
maximum should not be considered as an initial reference time of the
tunneling process.  Nevertheless, for electric field strengths well
below the threshold regime the electron exits the barrier after the
instant of electric field maximum.  For such strengths a nonvanishing
delay $\tau_\mathrm{exit}$ between the electric field maximum and the
emergence of the electron behind the tunneling barrier exists as a
signature of the time spent under the barrier.  The actual time span
that the electron has spent in the classically forbidden region,
however, will be larger than $\tau_\mathrm{exit}$.  Vanishing or
negative time delays $\tau_\mathrm{exit}$ should not be interpreted as
instantaneous tunneling.

Under the barrier as well at the tunneling exit, the electron has a
nonzero velocity in electric field direction.  After the tunneling exit,
the quantum trajectory, which is induced by the wave-function's
probability current, agrees well with a trajectory as given by
classical equations of motion when the electron's initial velocity at
the tunneling exit is taken into account properly.

\acknowledgments

We gratefully thank Enderalp Yakaboylu for valuable discussions.

\appendix*

\section{The Coulomb problem and the Strak effect in~two~dimensions}
\label{sec:appendix}

The eigen-equation of the two-dimensional Coulomb problem
\cite{Tanaka:Kobashi:etal:1987:LoSurdo-Stark_effect,Yang:Guo:etal:1991:Analytic_solution}
with some additional homogeneous electric field of strength
$\mathcal{E}$ is given in Cartesian coordinates $x$ and $y$ and
employing atomic units by
\begin{equation}
  \label{eq:Colomb_eigen_2d}
  \left(
    -\frac{1}{2}\frac{\partial^2}{\partial x^2}
    -\frac{1}{2}\frac{\partial^2}{\partial y^2}
    -\frac{1}{\sqrt{x^2 + y^2}} 
    -\mathcal{E} x
  \right)\Psi(x, y) = E \Psi(x, y) \,,
\end{equation}
where $\Psi(x, y)$ denotes an eigen-function with energy $E$.  This
eigen-equation separates in parabolic coordinates $\xi$ and $\eta$
\cite{Yang:Guo:etal:1991:Analytic_solution,Aquilanti:Cavalli:etal:1997:d-dimensional_hydrogen,Slen:Nepstad:etal:2007:N_-dimensional},
which are related to the Cartesian coordinates $x$ and $y$ via
Eq.~\eqref{eq:x_y_xi_eta}.  This coordinate system is particularly
useful because here the two-dimensional Schr{\"o}dinger equation can be
separated into two one-dimensional Schr{\"o}dinger equations, which allows
us to define a one-dimensional tunneling barrier.  The calculations in
this section follow Ref.~\cite{landau1977quantum}, where the
three-dimensional Coulomb problem is considered in a similar way.

The Laplacian becomes in the new coordinate system
\begin{equation}
  \label{eq:parlap}
  \frac{\partial^2}{\partial x^2} + 
  \frac{\partial^2}{\partial y^2} = 
  \frac{1}{\xi+\eta}\left(
    4\xi\frac{\partial^2}{\partial \xi^2} +
    2\frac{\partial}{\partial\xi} +
    4\eta\frac{\partial^2}{\partial\eta^2} + 
    2\frac{\partial}{\partial\eta}
  \right) \,. 
\end{equation}
Expressing \eqref{eq:Colomb_eigen_2d} in parabolic coordinates yields
after some algebraic transformations
\begin{multline}
  \label{eq:Colomb_eigen_2d_parabolic}
  \left(
     \xi\frac{\partial^2}{\partial \xi^2} +
     \frac12\frac{\partial}{\partial\xi} +
     \frac E2\xi + \frac{\mathcal{E}}{4}\xi^2 
  \right)\Psi(\xi, \eta) + \\
  \left(
    \eta\frac{\partial^2}{\partial \eta^2} +
    \frac12\frac{\partial}{\partial\eta} +
    \frac E2\eta - \frac{\mathcal{E}}{4}\eta^2 
  \right)\Psi(\xi, \eta) = -\Psi(\xi, \eta) \,.
\end{multline}
Substituting the ansatz $\Psi(\xi, \eta)=f_1(\xi)f_2(\eta)$ into the
last equation end separating the variables $\xi$ and $\eta$ we obtain
\begin{subequations}
  \begin{align}
    \label{eq:sep_xi}
    \left(
      \xi\frac{\partial^2}{\partial \xi^2} +
      \frac12\frac{\partial}{\partial\xi} +
      \frac E2\xi + \frac{\mathcal{E}}{4}\xi^2 + \beta_1
    \right)f_1(\xi) & = 0\,,\\
    \label{eq:sep_eta}
    \left(
      \eta\frac{\partial^2}{\partial \eta^2} +
      \frac12\frac{\partial}{\partial\eta} +
      \frac E2\eta - \frac{\mathcal{E}}{4}\eta^2 + \beta_2
    \right)f_2(\eta) & = 0\,,
  \end{align}
\end{subequations}
where the separation constants $\beta_1$ and  $\beta_2$ are related by
\begin{equation}
  \label{eq:sep_cond}
  \beta_1 + \beta_2 = 1 \,.
\end{equation}

The tunneling barriers are obtained by substituting
$f_1(\xi)=g_1(\xi)/\xi^{1/4}$ and $f_2(\eta)=g_2(\eta)/\eta^{1/4}$,
which gives the equations for the new functions as
\begin{subequations}
  \begin{align}
    -\frac{1}{2}\frac{\partial^2g_1(\xi)}{\partial\xi^2} + 
    \left(
      -\frac{3}{32\xi^2} - \frac{\beta_1}{2\xi} - \frac{\xi}{8}\mathcal{E}
    \right)g_1(\xi) & = \frac{E}{4}g_1(\xi)\,,\\
    -\frac{1}{2}\frac{\partial^2g_2(\eta)}{\partial\eta^2} + 
    \left(
      -\frac{3}{32\eta^2} - \frac{\beta_2}{2\eta} + \frac{\eta}{8}\mathcal{E}
    \right)g_2(\eta) & = \frac{E}{4}g_2(\eta)\,.
  \end{align}
\end{subequations}
These two equations represent Schr{\"o}dinger-type eigen-equations with the
potentials
\begin{subequations}
  \begin{align}
    \label{eq:tunnel_pot}
    V_1(\xi) & =
    -\frac{3}{32\xi^2} - \frac{\beta_1}{2\xi} - \frac{\xi}{8}\mathcal{E}\,,\\
    V_2(\eta) & = -\frac{3}{32\eta^2} - \frac{\beta_2}{2\eta} +
    \frac{\eta}{8}\mathcal{E} \,,
  \end{align}
\end{subequations}
and the energy $E/4$.

For a vanishing external electric field, i.\,e., $\mathcal{E}=0$,
Eqs.~\eqref{eq:sep_xi} and \eqref{eq:sep_eta} are identical.
Introducing the variable transformation $\xi=\eta=\zeta/\sqrt{-2E}$
gives in this case
\begin{equation}
  \left(
    \zeta\frac{\partial^2}{\partial \zeta^2} +
    \frac12\frac{\partial}{\partial\zeta} +
    \frac{\beta_{1,2}}{\sqrt{-2E}} - \frac{\zeta}{4}
  \right) f_{1,2}(\zeta) = 0 \,.
\end{equation}
Making the ansatz $f_{1,2}(\zeta)=\exp(-\zeta/2)g_{1,2}(\zeta)$ 
this differential equation yields the equation
\begin{equation}
  \left(
    \zeta\frac{\partial^2}{\partial \zeta^2} +
    \left(\frac12-\zeta\right)\frac{\partial}{\partial\zeta} -
    \left(\frac{1}{4} - \frac{\beta_{1,2}}{\sqrt{-2E}}\right)
  \right) g_{1,2}(\zeta) = 0 
\end{equation}
for $g_{1,2}(\zeta)$, which can be identified as the confluent
hypergeometric equation \cite{Arfken:Weber:2013:Mathematical_Methods}
with 
\begin{align}
  a & = \frac{1}{4} - \frac{\beta_{1,2}}{\sqrt{-2E}}\,, &
  b & = \frac12 \,.
\end{align}
The nonsingular solution of the confluent hypergeometric equation is
usually called confluent hypergeometric function and denoted by
${}_1F_1(a; b; \zeta)$ or $M(a; b; \zeta)$.  Thus we have found the
solution
\begin{equation}
  f_{1}(\xi) \sim \exp\left(-\frac{\xi}{2}\sqrt{-2E}\right) 
  M\left(
    \frac{1}{4} - \frac{\beta_{1}}{\sqrt{-2E}}; \frac12; \xi\sqrt{-2E}
  \right) 
\end{equation}
and similarly for $f_{2}(\eta)$.  These functions $f_{1}(\xi)$ and
$f_{2}(\eta)$ can be normalized only if the first argument of the
confluent hypergeometric function is a negative integer or zero.  In
this case the confluent hypergeometric function coincides (up to
normalization) with the associated Laguerre polynomials.  Thus, 
 the quantization conditions
  \begin{align}
    \frac{1}{4} - \frac{\beta_{1,2}}{\sqrt{-2E}} & = - n_{1,2}
  \end{align}
with $n_{1,2}=0,1,2,\dots$ will have to be fulfilled.  Together with the
relation \eqref{eq:sep_cond} the quantization conditions yield
\begin{align}
  \label{eq:E_2d_Coulomb}
  E & = -\frac{1}{2(n_1+n_2+1/2)^2} \,\,, \\
  \beta_{1,2} & = \left(n_{1,2}+\frac14\right)\sqrt{-2E}
  \,\,.
\end{align}
Normalizing $f_{1}(\xi)$ and $f_{2}(\eta)$ to unity finally gives the
bound eigen-states of the two-dimensional Coulomb problem
$\Psi_{n_1,n_2}(\xi, \eta)=f_{1;E,n_1}(\xi)f_{2;E,n_2}(\eta)$ with
\begin{subequations}
  \begin{align}
    f_{1;E,n_1}(\xi) & = \sqrt{\frac{\sqrt{-2E}}{1+4n_1}}
    \exp\left(-\frac{\xi}{2}\sqrt{-2E}\right)
    M\left(-n_1; \frac12; \xi\sqrt{-2E}\right) \,, \\
    f_{2;E,n_2}(\eta) & = \sqrt{\frac{\sqrt{-2E}}{1+4n_2}}
    \exp\left(-\frac{\eta}{2}\sqrt{-2E}\right) 
    M\left(-n_2; \frac12; \eta\sqrt{-2E}\right) \,,
  \end{align}
\end{subequations}
and the energy $E$ given by \eqref{eq:E_2d_Coulomb}.

\begin{figure}
  \centering
  \includegraphics[scale=0.45]{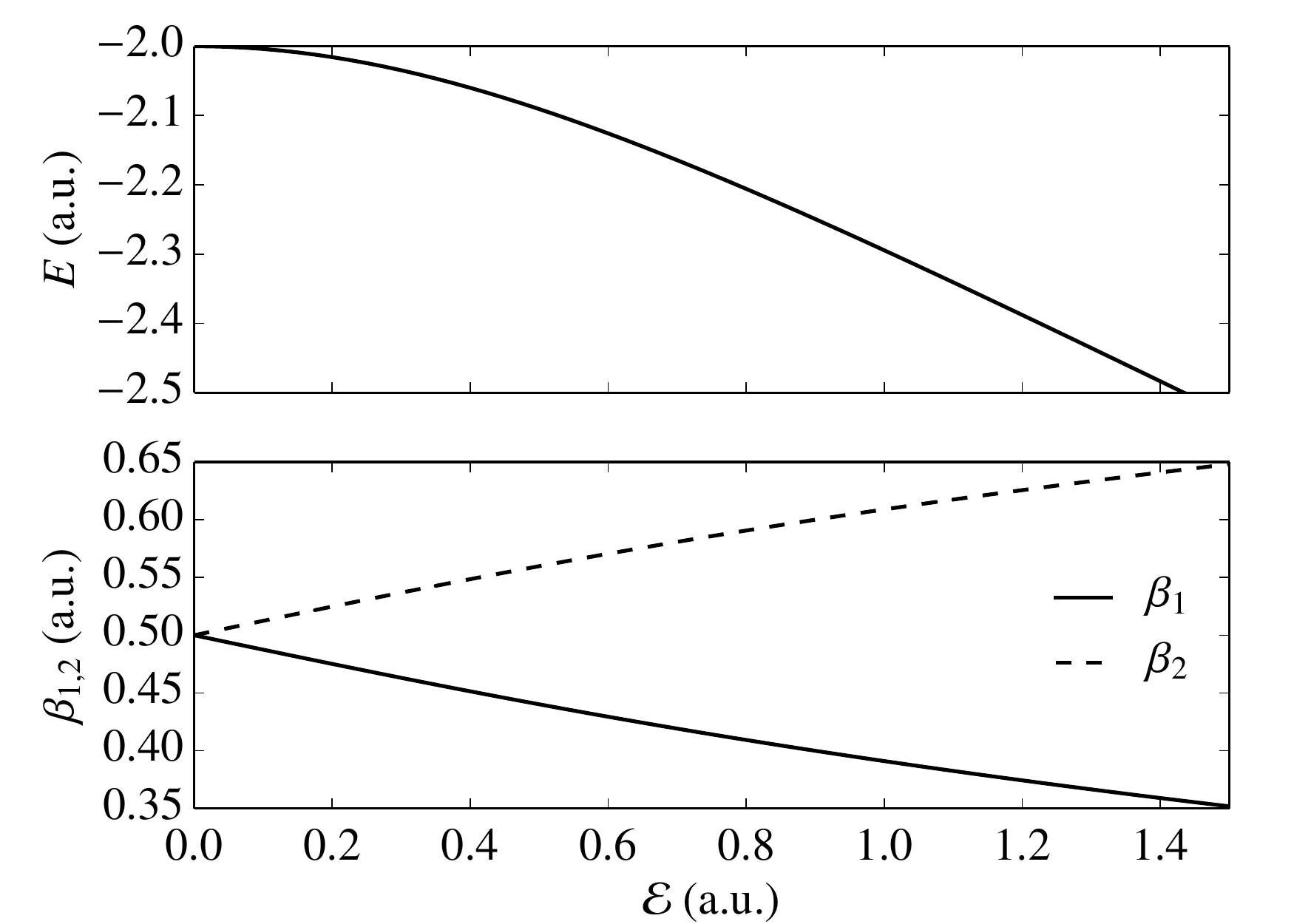}
  \caption{The ground state energy $E$ of the two-dimensional Coulomb
    problem with an external electric field and the separation
    parameters $\beta_{1}$ and $\beta_{2}$ as functions of the electric
    field strength $\mathcal{E}$.}
  \label{fig:perturbation_theory}
\end{figure}

Taking into account the Stark effect, i.\,,e., $\mathcal{E}>0$, leads to
a modification of the eigen-states, of the eigen-energies $E$, as well
as of the separation constants $\beta_1$ and $\beta_2$.  In this way,
the tunneling potential \eqref{eq:tunnel_pot} also changes.  The
resulting values for $\beta_1$, $\beta_2$, and $E$ can be calculated via
perturbation theory \cite{landau1977quantum}.  In case of the ground
state with $n_1=n_2=0$, we get in second order
\begin{equation}
  \label{eq:beta_12}
  \beta_{1,2} = \beta_{1,2}^{(0)} + \beta_{1,2}^{(1)} + \beta_{1,2}^{(2)} 
\end{equation}
with
\allowdisplaybreaks
\begin{subequations}
  \begin{align}
    \beta_{1,2}^{(0)} & = \sqrt{-E/8} \,, \\
    \beta_{1,2}^{(1)} & = \int f_{1,2;E,0}(\xi) \left(
      \mp\frac{\xi}{4}\mathcal{E} \right)
    f_{1,2;E,0}(\xi) \,\D\xi = \mp\frac{\mathcal{E}}{4E} \,, \\
    \nonumber
    \beta_{1,2}^{(2)} & = \frac{1}{\sqrt{-2E}}\times{}\\
    \nonumber
    & \qquad \sum_{n=1}^\infty
    \frac{1}{\frac14-\left(n+\frac14\right)}\left|
      \int f_{1,2;E,n}(\xi) \left(
        \mp\frac{\xi}{4}\mathcal{E} \right)
      f_{1,2;E,0}(\xi) \,\D\xi 
    \right|^2 \\
    & \approx 0.2004642410 \sqrt{-2E}\frac{\mathcal{E}^2}{E^3}\,.
  \end{align}
\end{subequations}
Equations~\eqref{eq:beta_12} and \eqref{eq:sep_cond} determine the
parameters $\beta_1$, $\beta_2$, and $E$ uniquely, and the results of a
numerical solution of these equations are shown in
Fig.~\ref{fig:perturbation_theory}.

\bibliography{virt_detector}

\end{document}